\documentstyle [12pt,eqsecnum,aps,amsfonts] {revtex}
\input epsf
\newcommand{\beq}{\begin{equation}}
\newcommand{\eeq}{\end{equation}}
\newcommand{\beqs}{\begin{eqnarray}}
\newcommand{\eeqs}{\end{eqnarray}}

\begin{document}

\draft

\baselineskip 6.0mm

\title{Study of the Potts Model on the Honeycomb and Triangular Lattices: 
Low-Temperature Series and Partition Function Zeros}

\vspace{4mm}

\author{
Heiko Feldmann$^{(a)}$\thanks{email: feldmann@insti.physics.sunysb.edu}, 
Anthony J. Guttmann$^{(b)}$\thanks{email: tonyg@maths.mu.oz.au}, 
Iwan Jensen$^{(b)}$\thanks{email: iwan@maths.mu.oz.au},
Robert Shrock$^{(a)}$\thanks{email: shrock@insti.physics.sunysb.edu}
\and Shan-Ho Tsai$^{(a)}$\thanks{email: tsai@insti.physics.sunysb.edu}}

\address{(a) \ Institute for Theoretical Physics \\
State University of New York \\ 
Stony Brook, N. Y. 11794-3840, USA} 

\address{(b) \ Department of Mathematics and Statistics \\
The University of Melbourne \\
Parkville, Vic 3052, Australia}

\maketitle

\vspace{4mm}

\begin{abstract}

   We present and analyze low-temperature series and complex-temperature 
partition function zeros for the $q$-state Potts model with $q=4$ on the 
honeycomb lattice and $q=3,4$ on the triangular lattice.  
A discussion is given as to how the locations of the 
singularities obtained from the series analysis correlate with the
complex-temperature phase boundary.  Extending our earlier work, we include a
similar discussion for the Potts model with $q=3$ on the honeycomb lattice and
with $q=3,4$ on the kagom\'e lattice. 
  
\end{abstract}


\vspace{10mm}

\pagestyle{empty}
\newpage

\pagestyle{plain}
\pagenumbering{arabic}
\renewcommand{\thefootnote}{\arabic{footnote}}
\setcounter{footnote}{0}

\section{Introduction}

  The 2D $q$--state Potts models \cite{potts,wurev} for various $q$ have been 
of interest as examples of different universality classes for phase transitions
and, for $q=3,4$, as models for the adsorption of gases on certain substrates.
Unlike the $q=2$ (Ising) case, however, for $q \ge 3$, the free energy has
never been calculated in closed form for arbitrary temperature.  It is thus of
continuing value to obtain further information about the 2D Potts model.  It
has long been recognized that a very powerful method for doing this is via the
calculation and analysis of series expansions for thermodynamic quantities such
as the specific heat, magnetization, and susceptibility \cite{grev}.  For
$q=2,3$, and 4, the respective 2D $q$-state Potts ferromagnets have continuous 
phase transitions, and the critical singularities and associated exponents 
are known exactly \cite{wurev,cns,cft}.  Recently,
two of us have calculated and analyzed long low-temperature series expansions
for the Potts model with $q=3$ on the honeycomb lattice and for the Potts
model with $q=3$ and $q=4$ on the kagom\'e lattice \cite{jge}.  
These have been used to make
very precise estimates of the respective critical points, to confirm a formula
for the honeycomb lattice and to strengthen a previous refutation of an old 
conjecture for the kagom\'e lattice.  
The other three authors have recently used a relation between
complex-temperature (CT) properties of the Potts model on a given lattice 
and physical properties of the Potts antiferromagnet (AF) on the dual lattice 
to rule out other conjectures \cite{hcl} and have calculated 
complex-temperature zeros of the partition function for these three 
cases of $q$ and lattice type \cite{p}.  The study of properties of
spin models with the magnetic field and temperature generalized to complex
values was pioneered by Yang and Lee \cite{yl} for the magnetic field and 
Fisher for the temperature \cite{mef}.   Some of the earliest work on CT
properties of spin models dealt with zeros of the partition function 
\cite{mef,kat,abe}.  Another
major reason for early interest in CT properties of spin models was the
fact that unphysical, CT singularities complicated the analysis of
low-temperature series expansions to get information about the location
and critical exponents of the physical phase transition \cite{dg}.  

   Here we shall present a unified study of the Potts model on the honeycomb
lattice for $q=4$ and on the triangular lattice for $q=3$ and $q=4$. For each
$q$ value and lattice type, our results include 
(i) long, low-temperature series for the specific
heat, spontaneous magnetization, and initial susceptibility  
derived using the finite-lattice method \cite{enting,enting96}, extended by 
noting the structure of the correction terms \cite{corr}; (ii) a calculation 
of the complex-temperature zeros and, from these, an inference about the CT 
phase boundary; and (iii) a discussion of how the positions of the physical 
and unphysical singularities extracted from the series analysis correlate 
with the CT phase boundary.  Since both the critical exponents and the location
of the paramagnetic-to-ferromagnetic phase transition are known exactly for
these models, we shall focus mainly on getting new information on
complex-temperature properties from the series and CT zeros.  Using the 
results of Refs. \cite{jge,hcl,p}, we shall also discuss subject (iii) for the 
Potts model with $q=3$ model on the honeycomb lattice and with $q=3$ and 4 on
the kagom\'e lattice.  It is useful to perform a unified analysis of this 
type because, aside from well-understood exceptions \cite{com}, the 
physical and complex-temperature singularities of the thermodynamic functions 
lie on the continuous locus of points ${\cal B}$ which serves as the 
boundaries of the complex-temperature phases \cite{cte}; consequently, an 
approximate knowledge (or exact knowledge, if available) of where
this boundary lies is of considerable help in checking which CT singularities
that one extracts from a series analysis are trustworthy and which are not.
This will be discussed further below.  
Note that low-temperature series on the honeycomb lattice correspond to
high-temperature series on the triangular lattice, and vice versa. 

\section{Model}

The (isotropic, nearest-neighbor) $q$-state Potts model at temperature $T$ 
on a lattice $\Lambda$ is defined by the partition function
\beq
Z = \sum_{ \{ \sigma_n \} } e^{-\beta {\cal H}}
\label{zfun}
\eeq
with the Hamiltonian
\beq
{\cal H} = J \sum_{\langle nn' \rangle}(1-\delta_{\sigma_n \sigma_{n'}})
+ H \sum_n(1-\delta_{0 \ \sigma_n }) 
\label{ham}
\eeq
where $\sigma_n=0,...,q-1$ are ${\Bbb Z}_q$-valued variables on each 
site $n \in \Lambda$, $\beta = (k_BT)^{-1}$, and $\langle n n' \rangle$
denotes pairs of nearest-neighbor sites. The symmetry group of the Potts 
Hamiltonian is the symmetric group on $q$ objects, $S_q$. 
We use the notation $K = \beta J$, 
\beq
a = z^{-1} = e^{K}
\label{a}
\eeq
\beq
x = \frac{e^K-1}{\sqrt{q}} \ . 
\label{x}
\eeq
and 
\beq
\mu = e^{-\beta H}
\label{mu}
\eeq
(The variable $z$ was denoted $u$ in Ref. \cite{jge}.)
The (reduced) free energy per site is denoted as 
$f = -\beta F = \lim_{N_s \to \infty} N_s^{-1} \ln Z$, where
$N_s$ denotes the number of sites in the lattice.
There are actually $q$ types of
external fields which one may define, favoring the respective values
$\sigma_n=0,..,q-1$; it suffices for our purposes to include only one.
The order parameter (magnetization) is defined to be
\beq
m = \frac{qM-1}{q-1}
\label{m}
\eeq
where $M = \langle \sigma \rangle =
\lim_{h \to 0} \partial f/\partial h$.  With this definition, $m=0$ in the
$S_q$-symmetric, disordered phase, and $m=1$ in the limit of saturated
ferromagnetic (FM) long-range order.  
Finally, the (reduced, initial) susceptibility is denoted as
$\bar\chi = \beta^{-1}\chi = \partial m/\partial h|_{h=0}$.
We consider the zero-field model, $H=0$, unless otherwise stated.  For $J >
0$ and the dimensionality of interest here, $d=2$, the $q$-state Potts model 
has a phase transition from the symmetric, high-temperature paramagnetic (PM) 
phase to a low-temperature phase involving spontaneous breaking of the 
$S_q$ symmetry and onset of ferromagnetic (FM) long-range order.  This
transition is continuous for $2 \le q \le 4$ and first-order for $q \ge 5$.  
As noted above, the model has the property of duality 
\cite{potts,wurev,kihara,kj}, which relates the partition function on a 
lattice $\Lambda$ with temperature parameter $x$ to another on the dual 
lattice with temperature parameter 
\beq
x_d \equiv {\cal D}(x) = \frac{1}{x} \ , \quad i.e.\quad
a_d \equiv {\cal D}(a) = \frac{a+q-1}{a-1} \ . 
\label{adual}
\eeq
Other exact results include formulas for the PM-FM transition temperature 
on the square, triangular, and honeycomb lattices \cite{potts,kj}, and 
calculations of the free energy at the phase transition temperature, and of the
related latent heat for $q \ge 5$ \cite{baxterf}. 
The case $J < 0$, i.e., the Potts antiferromagnet (AF), has also been of
interest because of its connection with graph colorings.  Depending on the 
type of lattice and the value of $q$, the antiferromagnetic model may have a 
low-temperature phase with AFM long-range order.  Alternatively, it may not
have any finite-temperature PM-AFM phase transition but instead may exhibit
nonzero ground state entropy.  For the Potts model on the honeycomb lattice,
the well-known $q=2$ (Ising) case \cite{ons,yang} 
falls into the former category, while the 
model with $q \ge 3$ falls into the latter category
\cite{sokal,p3afhc} with nonzero ground state entropy 
\cite{p3afhc,kewser,w,w3}.  
Reviews of the model include Refs. \cite{wurev,martinbook}.

  For the $q$-state Potts model, from duality and a star-triangle relation,
together with a plausible assumption of a single transition, one can derive 
algebraic equations that yield the PM-FM critical points for the honeycomb (hc)
and triangular (t) lattices \cite{kj}.  The equation for the honeycomb 
lattice is 
\beq
x^3 -3x-\sqrt{q}=0 \ , \quad i.e., \quad a^3-3a^2-3(q-1)a-q^2+3q-1=0 \quad 
{\rm (honeycomb)} 
\label{hceqa}
\eeq
and, as follows from eq. (\ref{adual}), the corresponding formula for the
triangular lattice is obtained by the replacement $x \to 1/x$:
\beq
\sqrt{q}x^3 +3x^2-1=0 \ , \quad i.e., \quad a^3-3a+2-q=0 \quad 
{\rm (triangular)} \ . 
\label{trieqa}
\eeq
It will be useful to have the explicit solutions for the cases studied here.
For $q=4$ on the honeycomb lattice, eq. (\ref{hceqa}) reduces to 
$(a-5)(a+1)^2=0$, yielding the PM-FM critical point 
\beq
a_{hc,PM-FM,q=4}=z_{hc,PM-FM,q=4}^{-1}=5 
\label{ahcc}
\eeq
together with a double root at the complex-temperature value
\beq
a_{hc,2,q=4}=z_{hc,2,q=4}^{-1}=-1
\label{ahcs}
\eeq
For $q=3$ on the triangular lattice, eq. (\ref{trieqa}) has the solutions 
\beq
a_{t,1,q=3} = a_{t,PM-FM,q=3} = \cos(2\pi/9)+\sqrt{3}\sin(2\pi/9)=1.879385...
\label{at1q3}
\eeq
i.e., $z_{t,PM-FM,q=3}=0.5320889...$, 
\beq
a_{t,2,q=3} = \cos(2\pi/9)-\sqrt{3}\sin(2\pi/9) = -0.347296...
\label{at2q3}
\eeq
and
\beq
a_{t,3,q=3} = -2\cos(2\pi/9) = -1.532089...
\label{at3q3}
\eeq
For $q=4$, eq. (\ref{trieqa}) reduces to $(a-2)(a+1)^2=0$, so that the physical
PM-FM critical point is given by
\beq
a_{t,1,q=4} = a_{t,PM-FM,q=4} = 2
\label{at1q4}
\eeq
and there is a double root at the CT value
\beq
a_{t,2,q=4} = -1 \ . 
\label{at2q4}
\eeq

\section{Series Expansions}

The low-temperature series expansion is based on perturbations from the
fully aligned ground state and is expressed in terms of the low-temperature
variable $z$ and the field variable $y=1-\mu$. Details of the 
methods can be found in Ref. \cite{jge}, so here it suffices to say that
in order to derive series in $z$ for the specific heat, magnetization
and susceptibility one need only calculate the expansion in $y$ 
to second order, i.e., 
\beq
Z = Z_{0}(z) + yZ_{1}(z) + y^{2}Z_{2}(z),
\label{zeq}
\eeq
where $Z_k(z)$ is a series in $z$ formed by collecting all
terms in the expansion of $Z$ containing factors of $y^k$.
We use the finite-lattice method \cite{enting96} to
approximate the infinite-lattice partition function $Z$ by a product of 
partition functions $Z_{m,n}$ on {\em finite} ($m\! \times \! n$) lattices,
with each $Z_{m,n}$ calculated by transfer matrix techniques.
As explained in Ref. \cite{jge}, this leads to a series in $z$ correct
to order $w_s(m-2)+m-1$, where $w_s$ is the maximal number of sites 
contained within the largest width $w$ of the rectangles, 
and $m$ is the number of nearest neighbors of each site.
The implementation of the algorithm on the honeycomb lattice \cite{jge}
has $w_s =2w$ and $m=3$. The triangular lattice is represented as
a square lattice with additional interactions along one of the diagonals,
and in this case $w_s = w$ and $m=6$. In addition, we make use of a recent
extension procedure discussed in Ref. \cite{corr}, which allows us to
calculate additional series terms. 

The extension procedure for the 4-state Potts model on the honeycomb 
lattice is the same as for the 3-state model \cite{jge}. 
For a given width the expansion is correct to order $2w+2$, and we 
calculated the series up to $w=12$. Next we look at the integer sequences 
$d_{s}(w)$ obtained by taking the difference between the expansions
obtained from successive widths $w$,
\beq
Z_{w+1}(z)-Z_{w}(z) = z^{2w+3} \sum_{s \geq 0}d_{s}(w) z^s.
\label{zweq}
\eeq
In this case the formulae for the correction terms are simply given 
by polynomials of order $2s+k$. We managed to find formulae for the first 
4 correction terms, which enabled us to calculate the series 
for the specific heat $C$, magnetization $m$, and susceptibility $\bar\chi$ 
to order 30. The resulting series for $m$, $\bar\chi$, and the (reduced)
specific heat $\bar C = C/(k_BK^2)$ are given in Table 1. 

The extension procedure for the triangular lattice is essentially the
same as for the honeycomb lattice. The only difference is that the order
of the polynomials is $s+k$. For a given width the expansion is correct to 
order $4w+5$, and we calculated the series up to $w=14$ for $q=3$ and up to
$w=12$ for $q=4$. We found formulae for the first 7 or 8 correction terms in
the case $q=3$ and the first 6 or 7 correction terms for $q=4$. The series 
were thus derived to order 69 (60) for the specific heat and magnetization 
and to order 68 (59) for the susceptibility in the case $q=3$ ($q=4$). 
The resulting series for $m$, $\bar\chi$, and the (reduced) specific heat 
$\bar C = C/(k_BK^2)$ are given in Tables 2 and 3. 

\section{Analysis of Series}

\subsection{Honeycomb Lattice, $q=4$}

   We have analyzed the series using dlog Pad\'e approximants (PA's) and 
differential approximants (DA's); for a general review of these methods, see
Ref. \cite{grev}.  We first comment on the physical PM-FM phase transition.  
The series yield a value for the critical point in excellent agreement with the
known value $z_{hc,PM-FM,q=4}=1/5$.  For example, the differential approximants
of the type $[L/M_0,M_1]$ with $L=1$ and $L=2$ to the specific heat series
yield $z_{hc,PM-FM,q=4}=0.19993(4)$ and 0.19991(5), while those for the
magnetization yield 0.19999(3) and 0.20005(7), respectively, with similarly
good agreement for other values of $L$ and for the approximants to the 
susceptibility.  Concerning the
critical exponents at this transition, the value $q=4$ is
the borderline between the interval $2 \le q \le 4$ where this transition is
second-order and the interval $q > 4$ where it is first order.  Related to
this, the $q=4$ 2D Potts model has the special feature that the thermodynamic
functions have strong confluent logarithmic corrections to their usual 
algebraic scaling forms \cite{cns} at the PM-FM transition (on any lattice).  
For example, the singularities in the specific heat and magnetization are 
$C_{sing} \sim |t|^{-2/3}(-\ln|t|)^{-1}$ 
for $t \to 0$, where $t=(T-T_c)/T_c$, and 
$M_{sing} \sim (-t)^{1/12}(-\ln|t|)^{-1/8}$ for $t \to 0^-$.  
Consequently, simple fits of the series to an algebraic
singularity without this confluent logarithmic correction are not expected to 
agree well with the known singularities.  Indeed, this was the general
experience in early series work, and the same is found for the longer series
here.  As an illustration, a naive fit to a simple algebraic singularity for
the specific heat would yield the value $\alpha' \sim 0.5$ rather than the
known value $\alpha'=2/3$.  Since these confluent singularities may also affect
singularities at complex-temperature points, it could be useful in future work,
as was noted earlier for the square-lattice model \cite{pfef}, to carry out 
a more sophisticated analysis of the series including these confluent 
singularities.  However, because our primary focus here is on obtaining new 
information on complex-temperature properties rather than reproducing exactly 
known results for the critical exponents of the physical PM-FM singularity, 
and because it is not known if the confluent logarithmic corrections do 
affect the CT singularities, we have not tried to include such logarithmic
factors in fits to the CT singularities.  

   Proceeding to CT singularities, we find evidence for one on the negative
real $z$ axis at 
\beq
z_{hc,\ell,q=4}=-0.33(1) \ , \quad i.e \quad a_{hc,\ell,q=4}=-3.0(1)
\label{zhcellq4}
\eeq
Here the subscript $\ell$ stands for ``leftmost'' singularity on the negative
real axis.  
We shall present below, as an application of the mapping discussed in
Ref. \cite{hcl}, an analytic derivation of the exact value 
$a_{hc,\ell,q=4}=-3$.  Clearly, the value extracted from the series analysis is
in excellent agreement with the exact determination.  By the mapping of
Ref. \cite{hcl}, it follows that the singularity in the specific heat at this
point $a_{hc,\ell,q=4}$, as approached from larger negative $a$, i.e. smaller
negative $z$, is the same as the singularity in the specific heat of the $q=4$
Potts antiferromagnet on the triangular lattice at the $T=0$ critical point 
as approached from finite temperature. 

   We also find evidence from the series analyses for at least one complex
conjugate (c. c.) pair of singularities.  One such pair is observed at 
\beq
z_{hc,cc1,q=4}, \ z_{hc,cc1,q=4}^* = 0.02(2) \pm 0.38(1)i
\label{zhccc1q4}
\eeq
The central values correspond to 
$a_{hc,cc1,q=4}, \ a_{hc,cc1,q=4}^* = 0.14 \pm 2.6i$.  As we shall show later,
this pair of singularities is consistent with lying on the complex-temperature
phase boundary ${\cal B}$. 

\subsection{Triangular Lattice, $q=3$}

The series yield values for the PM-FM critical point in excellent
agreement with the exactly known expression, eq. (\ref{at1q3}).  For 
example, the first-order DA's of the form $[L/M_0,M_1]$ with
$L=1$ for the free energy yield $z_{t,PM-FM,q=3}=0.532095(85)$, 
in complete agreement, to within the uncertainty, with the known value given 
by eq. (\ref{at1q3}). For reference, the thermal and field exponents for the 2D
$q=3$ Potts model are $y_t=6/5$ and $y_h=28/15$, so that the critical exponents
for the specific heat, magnetization, and susceptibility are 
$\alpha=\alpha'=1/3$, $\beta=1/9$, and $\gamma=\gamma'=13/9=1.444...$ 
\cite{wurev,cft}.  The above approximants yield the exponent
$\alpha'=0.331(27)$, again in agreement with the known value.  Similar
statements apply to the magnetization and susceptibility. 

   Concerning complex-temperature singularities, the series for $m$ and 
$\bar\chi$ indicate a singularity on the negative real
axis, at $z_{t,-,q=3} \simeq -0.71$ and $z_{t,-,q=3} \simeq -0.65$.  If we 
assume that this is, as it should be, the same singularity, and average the 
positions, we get
\beq
z_{t,-,q=3} \equiv z_{t,\ell,q=3} = -0.68(5)
\label{ztellq3ser}
\eeq
or equivalently,
\beq
a_{t,\ell,q=3}=-1.47(11)
\label{atellq3ser}
\eeq
where the numbers in parentheses are our estimates of the theoretical
uncertainties.  We observe that our numerical determination in eq. 
(\ref{atellq3ser}) is consistent, to within the uncertainty, with being equal 
to the value given by the root in eq. (\ref{at3q3}), whence our use of the
symbol $a_{t,\ell,q=3}$ in eq. (\ref{atellq3ser}).

   We find a complex-conjugate pair of singularities at
\beq
z_{t,e,q=3}, \ z_{t,e,q=3}^* = 0.0209(1) \pm 0.531(1)i
\label{zeq3ser}
\eeq
 From our analysis of the
respective series, we infer the following values of singular exponents at the
points (\ref{zeq3ser}): 
\beq
(\alpha', \ \beta, \ \gamma')_{z_{t,e,q=3}} = (1.19(1), \ -0.18(1), \ 1.17(1))
\label{expzeq3ser}
\eeq
The central values in eq. (\ref{zeq3ser}) correspond to
\beq
a_{t,e,q=3}, \ a_{t,e,q=3}^* = 0.0740 \pm 1.88i
\label{aeq3ser}
\eeq
 From the CT zeros to be discussed below, we see clearly that the c. c. members
of this pair are endpoints of arcs of CT zeros of $Z$, corresponding to
continuous arcs of singularities of the free energy in the thermodynamic
limit. (This type of correspondence with endpoints on ${\cal B}$ is indicated 
by the subscripts $e$ here and in other cases below.)  In passing, we observe 
that the exponent values in eq. (\ref{expzeq3ser}) are not too different from  
the respective exponents obtained from the series analysis of Ref. \cite{ghs}
the singularities $u_s = -0.301939(5) \pm 0.3787735(5)i$ in the 2D spin-1 
Ising model on the square lattice, namely \cite{ghs} 
$(\alpha', \ \beta, \ \gamma') = (1.1693(3), \ -0.1690(2), \ 
1.1692(2))$.  The c.c. pair of points $u_s$ is analogous to the pair in 
eq. (\ref{aeq3ser}) because the members of this pair 
were shown \cite{hs} to be endpoints of 
arcs of CT zeros protruding into the complex-temperature extension of the FM 
phase of the spin-1 square-lattice Ising model.   We also observe that the
values of both these sets of exponents are consistent with the equality 
$\alpha' = \gamma'$.  However, we already know that such an equality, even if
it were to hold for these cases, is not a general result for singular exponents
at endpoints of arcs of a CT boundary ${\cal B}$ protruding into the
complex-temperature extension of the FM phases for a spin model.  A
counterexample is provided by the (isotropic, spin 1/2) Ising model on the
triangular lattice.  In this case, one can determine the complex-temperature
 phase diagram exactly, and ${\cal B}$ consists of the union of the unit circle
 $|u+1/3|=2/3$ and the semi-infinite line segment $-\infty \le u \le -1/3$
\cite{chitri}, where $u=z^2$.  Thus, in this case there is an exactly known
 analogue to the arc endpoints, viz., the endpoint at $u_e=-1/3$ (where the
 subscript $e$ denotes ``endpoint'') of the line
 segment protruding into the complex-temperature extension of the FM phase.  An
 analysis of low-temperature series \cite{g75} had earlier yielded the 
inference that $\gamma_e' = 5/4$, while exact results 
\cite{chitri} yielded $\alpha_e' = 1$ (and $\beta_e=-1/8$), so that 
$\alpha_e' \ne \gamma_e'$. 

We find a second c. c. pair at
\beq
z_{t,e',q=3},z_{t,e',q=3}^* = -0.515(3) \pm 0.322(3)i
\label{zfq3ser}
\eeq
with exponents $(\alpha', \ \beta, \ \gamma') = (1.2(1), \ -0.25(10), \ 
1.2(1))$.  The central values in eq. (\ref{zfq3ser}) correspond to
\beq
a_{t,e',q=3},a_{t,e',q=3}^* = -1.40 \pm 0.873i
\label{afq3ser}
\eeq
This pair is consistent with lying on the CT phase boundary, as will be
discussed below.  It should be noted that we would not expect the 
low-temperature series to be sensitive to the third root, $a_{t,2,q=3}$, of 
eq. (\ref{trieqa}), since this root is masked by the nearer singularity 
$a_{t,3,q=3}$
(that is, $z_{t,3,q=3}=-0.652704...$ is closer to the origin in the $z$ plane
than $z_{t,2,q=3}=-2.879385..$).  

\subsection{Triangular Lattice, $q=4$} 

   For the physical PM-FM critical point of the $q=4$ Potts model on the
triangular lattice, the discussion that we gave above for the honeycomb 
lattice applies; that is to say, the position of the physical singularity is
well approximated, but the critical exponents are not, due to the presence of
confluent logarithms.  For complex-temperature properties, we first note that
the series do not give a firm indication of a singularity on the negative 
real axis.  We find a complex-conjugate pair of singularities at
\beq
z_{t,e,q=4}, \ z_{t,e,q=4}^* = 0.0304(2) \pm 0.498(2)i
\label{zeq4ser}
\eeq
We have also studied the exponents at this pair of singularities.  If one
assumes that there are no strong confluent singularities present, such as the
logarithms that are present at the physical critical point, then from our
series analysis we extract the following values, with their quoted
uncertainties:
\beq
(\alpha', \ \beta, \ \gamma')_{z_{t,e,q=4}} = (1.18(2), \ -0.17(2), \ 1.20(2))
\label{expzeq4ser}
\eeq
However, we caution that it is not known whether strong confluent 
singularities are present at the points (\ref{zeq4ser}), and if they are, then
the values in eq. (\ref{expzeq4ser}) would have a lower degree of reliability. 
The central values in eq. (\ref{zeq4ser}) correspond to
\beq
a_{t,e,q=4}, \ a_{t,e,q=4}^* = 0.122 \pm 2.00i
\label{aeq4ser}
\eeq
As in the $q=3$ case, from the CT zeros to be presented below,
we see clearly that the c. c. members
of this pair of singularities are endpoints of arcs of CT zeros of $Z$.

We find a second c. c. pair at
\beq
z_{t,e',q=4}, \ z_{t,e',q=4}^* = -0.461(5) \pm 0.281(5)i
\label{zfq4ser}
\eeq
The central values correspond to
\beq
a_{t,e',q=4}, \ a_{t,e',q=4}^* = -1.58 \pm 0.964i
\label{afq4ser}
\eeq
Again, this pair can be associated with endpoints of arcs of zeros, as is
especially clear from Fig. \ref{q4hexpfaDual}.  There is also some sign of
another pair of singularities in the vicinity of $z \simeq -0.2 \pm 0.6i$,
corresponding to $a \simeq -0.5 \pm 1.5i$. The members of this c. c. pair are
consistent with lying on the CT phase boundary.  It is possible that there are
also other c. c. pairs of singularities.

\section{Complex-Temperature Zeros}

\subsection{General} 

   The (zero-field) Potts model partition function $Z$ for a finite lattice 
is, up to a possible prefactor, a polynomial in the Boltzmann weight $a$. 
We calculate this polynomial by standard transfer matrix methods.  From this, 
we then compute the zeros.  
In the thermodynamic limit, via a coalescence of zeros, there forms a 
continuous locus ${\cal B}$ of points where the free energy is nonanalytic. As
was noted, this locus serves as the union of boundaries of the various 
complex-temperature phases \cite{cte} and, aside from well-understood
exceptions \cite{com}, the 
CT singularities of thermodynamic functions occur on the continuous locus of
points ${\cal B}$ where the free energy is nonanalytic, since it is analytic
in the interior of physical phases and their complex-temperature extensions. 
Thus, calculations of CT zeros on sufficiently large finite lattices yield
useful information on the CT phase diagram in the thermodynamic limit.  
Hence, when investigating CT singularities, it is useful
to do so in conjunction with a calculation of the CT zeros of the partition
function to infer the approximate location of the CT phase boundary
${\cal B}$.  

   To illustrate this, let us return briefly to the $q=2$ Ising special case 
of the
Potts model, for which both the free energy \cite{ons} and the magnetization
\cite{yang} are known exactly.  We recall that the expression for the 
spontaneous magnetization is \cite{yang}
\beq
M=\frac{(1+u)^{1/4}(1-6u+u^2)^{1/8}}{(1-u)^{1/2}}
\label{m_ising}
\eeq
where $u=z^2$ in the FM phase and
the complex-temperature extension of it (with $M=0$ elsewhere). 
Let us pretend that we did not know the exact free energy or magnetization, but
that we had a low-temperature (small-$u$) series for $M$ and analyzed it using
dlog Pad\'e approximants.  We would find singularities at the following four
points: (i) $u_{PM-FM}=3-2\sqrt{2}$, the physical PM-FM phase transition 
point; (ii) $u_{PM-AFM}=u_c^{-1}=3+2\sqrt{2}$, the PM-AFM transition point; 
(iii) $u=-1$; and (iv) $u=1$.   This shows the value and importance of
analyzing CT singularities of thermodynamic functions from series expansions;
these can give one deeper knowledge of the exact functions.  Indeed, in this
illustrative example, a dlog Pad\'e analysis of the series for $M$ would enable
one to reconstruct the exact analytic expression for this quantity.  The
knowledge of the CT zeros and corresponding CT phase diagram give complementary
information, in particular, information on which of the singularities found
from the series analysis occur in the true thermodynamic function.  Thus, from 
calculations of CT zeros for finite lattices, we could determine the 
approximate CT phase boundaries, 
and, in particular, the CT extension of the FM phase.  We would then infer that
this (CT) FM phase does not include the point $3+2\sqrt{2}$, so that the second
apparent singularity extracted from the series does not occur in the true
function $M$, since the low-temperature series only apply in the physical FM
phase and its complex-temperature extension.  
Even without the CT zeros, in this case, we would also know that
the apparent singularity at $u=1$ does not occur in the true $M$, since $M$ 
is certainly zero, and all thermodynamic functions are analytic, at the 
infinite-temperature point $u=1$ (and since this point is clearly in the PM
phase, the low-temperature series are again not applicable in its vicinity).
This example thus illustrates the value of both the study of CT singularities
from series expansions and CT zeros of the partition function.  
Here, of the four apparent singularities extracted from the series, 
only two, namely the physical critical point (i) and the CT singularity (iii) 
are true singularities of $M$, since the others occur in regions outside the CT
extension of the FM phase where the series applies.  In this exactly solved
case, these results are obvious, but the lesson holds more generally and
illustrates the usefulness of having at least approximate knowledge of the CT
phase boundary of a given model when analyzing series expansions to obtain
locations of CT singularities.  Note that all of the CT singularities occur on
the locus of points ${\cal B}$ where the free energy is nonanalytic, which in
this case is a lima\c{c}on of Pascal (given by eqs. (2.17) and (2.18) in
Ref. \cite{chisq}, the image in the $u$ plane of the circles \cite{mef}
$|z \pm 1| = \sqrt{2}$ in the $z$ plane.) This is also a general feature (with
the exception noted in Theorem 6 of Ref. \cite{cmo}) and constitutes another
reason for the value that a knowledge of CT zeros and the corresponding locus 
${\cal B}$ have for series analyses and vice versa. 

  As a technical remark, we note that the problem of calculating the zeros of
the partition function for large lattices is a challenging one, since the
degree of the polynomial is equal to the number of bonds, 
$N_b = (\Delta/2)N_s$, where $\Delta$ is the coordination number, $N_s$ is the
number of sites, and there is a very large range in the
sizes of the coefficients, from $q$ for the highest-degree term $a^{N_b}$ to
exponentially large values for intermediate terms.  The latter property is
obvious from the fact that for $K=0$, i.e., $a=1$, the sum of the 
coefficients in $Z$ is $q^{N_s}$.  A general property of the CT phase
boundary for any lattice and $q$ value is invariance under complex-conjugation:
${\cal B} \to {\cal B}$ as $a \to a^*$.  For previous calculations of CT 
zeros for the Potts model on the triangular and square lattices, see 
\cite{mm,wuz,pfef,martinbook}. 

   The complex-temperature zeros are equivalently plotted in the complex $a$ 
or $z$ plane; we shall plot them in the $a$ plane because the resultant 
figures are somewhat more compact and because this maintains conformity with 
the plots for the square lattice,
where the $Re(a) > 0$ part of the phase boundary is very simple (either exactly
or approximately part of a circle, depending on boundary conditions). 

   In making inferences about the CT phase boundary ${\cal B}$ in the
thermodynamic limit from calculations on CT zeros on finite lattices, it is
important to get an idea of the effects of different boundary conditions and
lattice sizes.  Accordingly, in Ref. \cite{p}, the authors performed 
calculations with three different boundary conditions and compared the
resultant CT zeros with the exactly known CT phase boundary for the $q=2$ 
Ising model on the honeycomb and kagom\'e lattices.  This also served as a 
check on the computer programs used. We shall use the same
three types of boundary conditions here, and we identify them next, following
the notation of Ref. \cite{p}.  

\subsection{Honeycomb Lattice, $q=4$}

\begin{figure}
\centering
\
\epsfxsize=8.5cm
\epsfbox{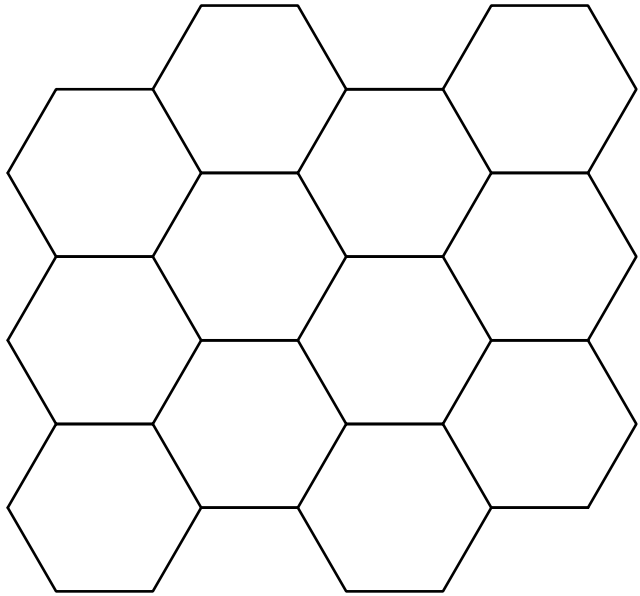}
\caption{Honeycomb lattice to illustrate our
conventions for indicating sizes.}
\label{lattice}
\end{figure}

 Before we start to present our results, we have to introduce our notation for
the sizes and orientations of the lattices. To indicate the size of a given
lattice, we count the number of hexagons.  As an illustration, the size of 
the honeycomb lattice in Fig. \ref{lattice} is $4 \times 3$ hexagons. 
The number of sites in a lattice is also dependent on the
boundary conditions: with
periodic boundary conditions in the horizontal direction for example,
the sites on the left and right are identified, while
with free boundary conditions they
are counted independently from each other.

Since we make use of duality in this work, we use lattices that have a
corresponding dual lattice. This excludes
lattices that are periodic in both directions, for the following reason:
duality relies on the fact that every closed polygon divides the
lattice into at least two regions.  However, a lattice with periodic boundary
conditions in both directions, and hence with toroidal geometry, has the
property that one can easily draw a closed contour that does not
divide the surface into two disjunct regions.  Since boundary effects are, in
general, best suppressed if one uses periodic boundary conditions in as many
directions as possible, we use boundary conditions that are periodic in one
direction and free in the other. Our notation for the boundary conditions
(BC's) is (fbc,pbc) for free and periodic BC's in the horizontal ($x$) and
vertical ($y$) directions, respectively (see Fig. \ref{lattice}), and
(pbc,fbc) for periodic and free BC's in the $x$ and $y$ directions.
Note that for the (fbc, pbc) choice, there is one site per hexagon at the
boundary with only two instead of the usual $\Delta=3$ bonds.
For the (pbc, fbc) BC's, there are two of these sites per boundary hexagon.
This motivated the formulation of a third kind of boundary condition \cite{p}:
starting from the (pbc, fbc) BC's, one adds bonds connecting the boundary sites
with fewer than three bounds so that all sites on the lattice have the same
coordination number $\Delta = 3$.  This type of boundary conditions is denoted
as (pbc,fbc)$_\Delta$.  In Ref. \cite{p}, a comparison was made with the
Ising case $q=2$ where the CT phase boundary is exactly known, and it was found
that (for the same lattice sizes as are used here) the CT zeros calculated with
all three types of boundary conditions tracked the exactly known CT phase 
boundary reasonably well.  In
particular, the (pbc,fbc)$_\Delta$ choice produced CT zeros with, in general,
the least scatter.  The (fbc,pbc) choice also exhibited the special feature
that a subset of zeros lay exactly on a certain circular arc comprising part of
${\cal B}$. 

\begin{figure}
\centering
\
\epsfxsize=8.5 cm
\epsfbox{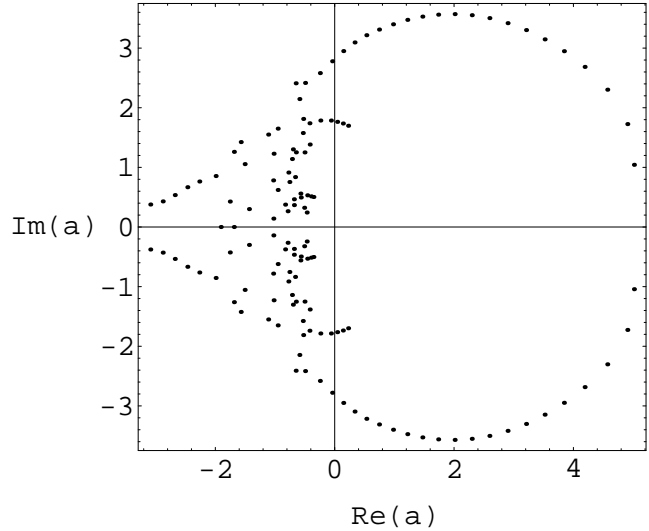}
\caption{CT zeros of $Z$ in the $a$ plane for the $q=4$ Potts model on a 
honeycomb lattice of size $7 \times 6$ hexagons and boundary conditions 
of type (fbc,pbc).}
\label{hex4fpa}
\end{figure}

\begin{figure}
\centering
\
\epsfxsize=8.5 cm
\epsfbox{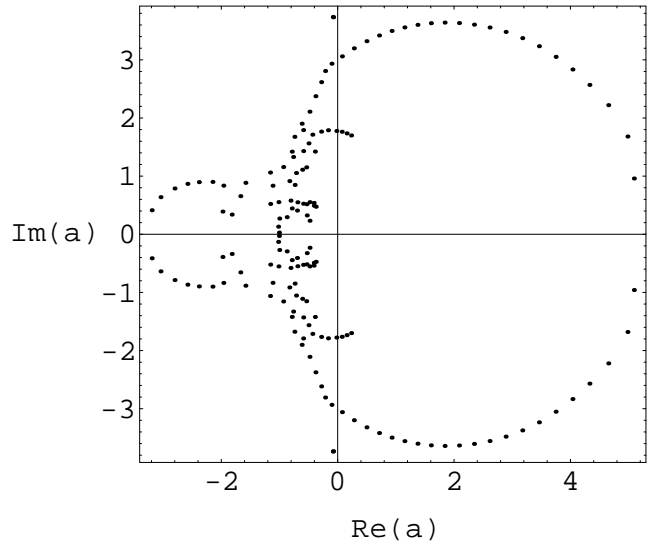}
\caption{CT zeros of $Z$ for the $q=4$ Potts model on a honeycomb lattice of 
size $8 \times 6$ hexagons and boundary conditions of type (pbc,fbc).}
\label{hex4pfa}
\end{figure}

\begin{figure}
\centering
\
\epsfxsize=10 cm
\epsfbox{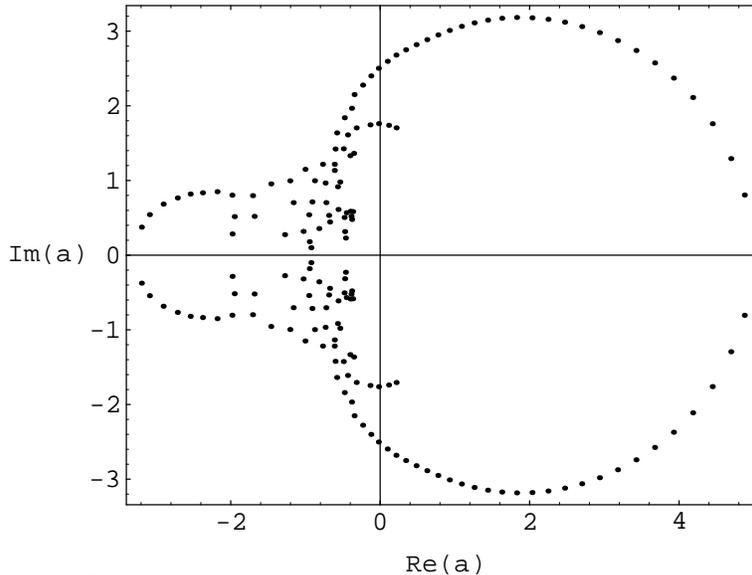}
\caption{CT zeros of $Z$ for the $q=4$ Potts model on a honeycomb lattice of
size $8 \times 6$ hexagons and boundary conditions of type (pbc,fbc)$_\Delta$.}
\label{hex4pfSpecial}
\end{figure}

   We show our calculations of the CT zeros of the $q=4$ Potts model on the
honeycomb lattice, using the above three types of boundary conditions, in
Figs. \ref{hex4fpa}, \ref{hex4pfa}, and \ref{hex4pfSpecial}.  
The zeros cross the positive real
$a$ axis only at one point, which is the PM-FM transition point; this value is
in good agreement with the exact result $a_c=5$ of eq. (\ref{ahcc}).  The zeros
thus exclude a PM-AFM transition and associated low-temperature phase with
antiferromagnetic long-range order, since such a transition would be 
represented by a curve of CT zeros crossing the real $a$ axis 
at some point in the interval $0 \le a < 1$.  
Concerning earlier work that bears on this, we note that a
recent Monte Carlo study of the $q=3$ Potts antiferromagnet on the honeycomb
lattice \cite{p3afhc} yielded evidence that that model has no symmetry-breaking
phase transition and thus is disordered at all temperatures, including $T=0$,
where it exhibits nonzero ground state entropy measured to be $S_0/k_B =
0.957$.  The latter value is close to an estimate \cite{p3afhc} from earlier
large-$q$ series \cite{kewser} and is bracketed closely by rigorous upper 
and lower bounds \cite{w3}.  Because increasing $q$ makes the spins
``floppier'', one expects that the Potts antiferromagnet on the honeycomb
lattice for $q \ge 4$ is similarly disordered at all temperatures, and, 
indeed, this has been rigorously proved \cite{sokal}.  

    A second remark is that the zeros are also consistent with the inference 
that a curve on the CT phase boundary ${\cal B}$ crosses the real axis at 
the value in eq. (\ref{ahcs}), $a=-1$. This
crossing is clearest with the (pbc,fbc) boundary conditions, shown in Fig. 
\ref{hex4pfa}.  Two other possible crossings occur at $a=-2.0(2)$ and 
$a=-0.47(5)$.  For the latter point we have another source of information,
using duality; if a crossing did occur at this point, it would be the closest,
on the left, to the origin of the $a$ plane and consequently its dual image
would be the leftmost crossing of the CT phase boundary of the $q=4$ Potts
model on the triangular lattice, in the $a_d$ plane. 
In the other cases of $q$ value and lattice type where such Potts model series
have been calculated and analyzed \cite{jge}, they have been able to locate the
leftmost singularity on the real $a$ axis (corresponding to the nearest
singularity left of the origin in the $z$ plane) with good accuracy.  However,
the analysis of the low-temperature series for the $q=4$ Potts model on the
triangular lattice does not yield very strong evidence for such a
singularity.  

   The leftmost crossing, in the $a$ plane, of the CT zeros for the $q=4$ 
Potts model on the honeycomb lattice, and hence of ${\cal B}$ in the
thermodynamic limit, is related by duality to physical properties of the $q=4$
Potts antiferromagnet on the dual, i.e. triangular, lattice; as discussed in
Ref. \cite{hcl}, the full temperature interval $0 \le a_d \le 1$ of the 
$q$-state Potts AF on this
dual lattice $\Lambda_d$ maps in a 1-1 manner, under duality, to the
complex-temperature interval $-\infty \le a \le -(q-1)$ of the Potts model on
the lattice $\Lambda$.  Now it has been argued \cite{baxter} that the Potts
AF on the triangular lattice has a zero-temperature critical point (see also
Ref. \cite{henley}, where a study of the closely related $T=0$ critical point 
of the $q=3$ Potts AF on the kagom\'e lattice is given).  Using the duality
connection \cite{hcl}, one then deduces that the leftmost crossing of 
${\cal B}$ for the $q=4$ Potts model on the honeycomb lattice is at 
\beq
a_\ell = -3
\label{aell}
\eeq
The CT zeros that we have calculated are consistent with this.  The slight 
flaring out of the zeros to the left of this point appears as a finite 
lattice-size effect. 

    As was the case with the $q=2$ Ising case and with $q=3$, we again observe
complex-conjugate arcs of zeros protruding into, and terminating in, 
the CT extension of the PM phase, ending at
\beq
a_{hc,e,q=4}, \ a_{hc,e,q=4}^* = 0.27(3) \pm 1.68(4)i
\label{ae}
\eeq
We recall that in the exactly known $q=2$ case, these arc endpoints occur at 
$a=e^{\pm \pi i/3}$; as discussed in \cite{p}, as $q$ increases, these arc
endpoints in the CT PM phase move to larger magnitudes $|a_e|$ and larger 
values of the angle $\theta=\arg(a_e)$. There also appears to be another c. c.
pair of arcs protruding into the CT PM phase, with endpoints at
\beq
a_{hc,e',q=4}, \ a_{hc,e',q=4}^* = -0.34(3) \pm 0.45(7)i
\label{aep}
\eeq

   A general observation is that all of the complex-temperature singularities
obtained from the analysis of the low-temperature series are consistent with
lying on the CT phase boundary ${\cal B}$.  This clearly 
includes the physical PM-FM critical point $a_{hc,PM-FM,q=4}=5$, the leftmost
crossing at $a_{hc,\ell,q=4}=-3$, and also the complex-conjugate pair given
in eq. (\ref{zhcellq4}).  We note that (i) the crossing at $a=-1$, and the 
c. c. pairs of arc endpoints in the PM phase in eqs. (ii) (\ref{ae}), and 
(iii) (\ref{aep}) are not expected to be seen with the low-temperature series 
because they are not contiguous with the complex-temperature extension of the
FM phase but rather are within a presumed O phase \cite{cte} 
for (i) and the CT PM phase for (ii) and (iii).  

\subsection{Triangular Lattice, $q=3$}

\begin{figure}
\centering
\
\epsfxsize=8.5 cm
\epsfbox{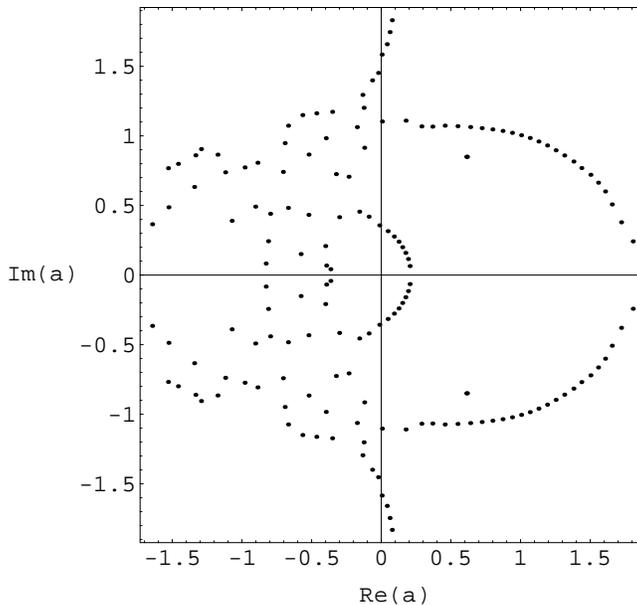}
\caption{CT zeros of $Z$ for the $q=3$ Potts model on the triangular lattice,
obtained via duality from a honeycomb lattice of size
$8 \times 6$ hexagons and (pbc,fbc) boundary conditions.}
\label{q3hexpfaDual}
\end{figure}

\begin{figure}
\centering
\
\epsfxsize=8.5 cm
\epsfbox{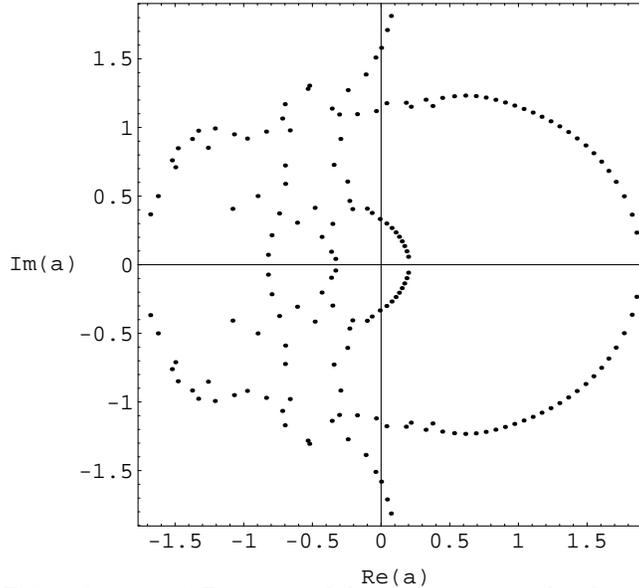}
\caption{CT zeros of $Z$ for the $q=3$ Potts model on the triangular lattice,
obtained via duality from results for a honeycomb lattice of size
$8 \times 6$ hexagons and (pbc,fbc)$_\Delta$ boundary conditions.}
\label{q3hexpfSpecaDual}
\end{figure}

   In Figs. \ref{q3hexpfaDual} and \ref{q3hexpfSpecaDual}, we present
calculations of CT zeros for the $q=3$ Potts model on the triangular lattice.
For consistency, we have
plotted all of our zeros in the $a$ plane; however, we note that when relating
zeros of $Z$ for the Potts model on one lattice $\Lambda$ to those on the dual
lattice $\Lambda_d$, the connection is simplest if one plots the zeros in the
$x$ plane, where $x$ was given in eq. (\ref{x}) since in this case the duality
transformation (\ref{adual}) just amounts to the inversion map $x \to 1/x$.
A comparison of these figures gives a quantitative indication of the effects
of different boundary conditions.  These effects are somewhat stronger for
$Re(a) < 0$ than $Re(a) > 0$.

One observes complex-conjugate arc endpoints protruding into, and ending in,
the (complex-temperature extension of the) FM phase at
\beq
a_e,a_e^* = 0.072(10) \pm 1.85(2)i
\label{aefromzeros}
\eeq
The points at which one can infer crossings of the zeros, and hence the CT
phase boundary ${\cal B}$ are (in order, proceeding from right to left along
the real $a$ axis) (1) the PM-FM critical point
$a_{PM-FM,q=3}=1.879...$ given by eq. (\ref{at1q3}); (2) the PM-AFM critical
point
\cite{grest,entingwu}, which has recently been measured by a Monte Carlo
simulation to high precision, $a_{PM-AFM,q=3}= 0.20309(3)$ \cite{adler};
(3) $a_{2,q=3}=-0.3473...$ in eq. (\ref{at2q3}), (4) $a=-0.82(3)$, and (5)
$a_{3,q=3}=-1.532...$ in eq. (\ref{at3q3}).  In Ref. \cite{mm}, it was
suggested that
points (2) and (4) were given by two of the roots of the equation
$a^3+6a^2+3a-1=0$, viz., $a=0.226682...$, $a=-0.8152075...$, while the third
root, $a=-5.411474...$ would be associated with the completion of the
complex-conjugate arcs of zeros (labelled as branches 6 in Ref. \cite{mm}) to
form a closed curve crossing the negative real $a$ axis at this root.
However, neither the early \cite{grest,entingwu} determinations nor the recent
high-precision determination \cite{adler} of $a_{PM-AFM}$ agrees with the value
$a=0.226682$, and the suggestion about the closing of the arcs to form a closed
curve crossing the negative real axis at $a=-5.411474$ has been refuted
\cite{hcl} since, by duality, it is equivalent to a finite-temperature phase
transition in the $q=3$ Potts antiferromagnet on the honeycomb lattice, which
is known not to occur \cite{p3afhc}.   The CT phase diagram \cite{cte}
in the $a$ plane for the $q=3$ Potts model on the triangular lattice thus
consists of a PM phase and an FM phase, with indications of at least one O
phase.

\subsection{Triangular Lattice, $q=4$}

\begin{figure}
\centering
\
\epsfxsize=8.5 cm
\epsfbox{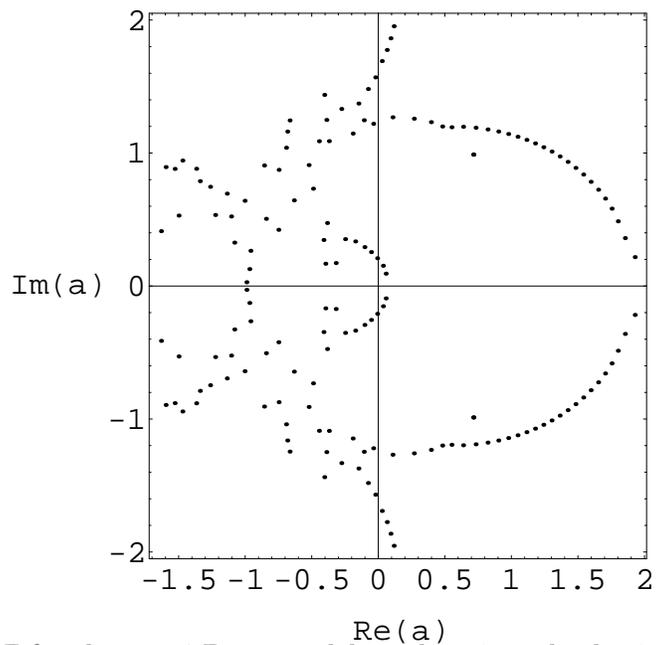}
\caption{CT zeros of $Z$ for the $q=4$ Potts model on the triangular lattice,
obtained via duality from a honeycomb lattice of size
$8 \times 6$ hexagons and (pbc,fbc) boundary conditions.}
\label{q4hexpfaDual}
\end{figure}

\begin{figure}
\centering
\
\epsfxsize=8.5 cm
\epsfbox{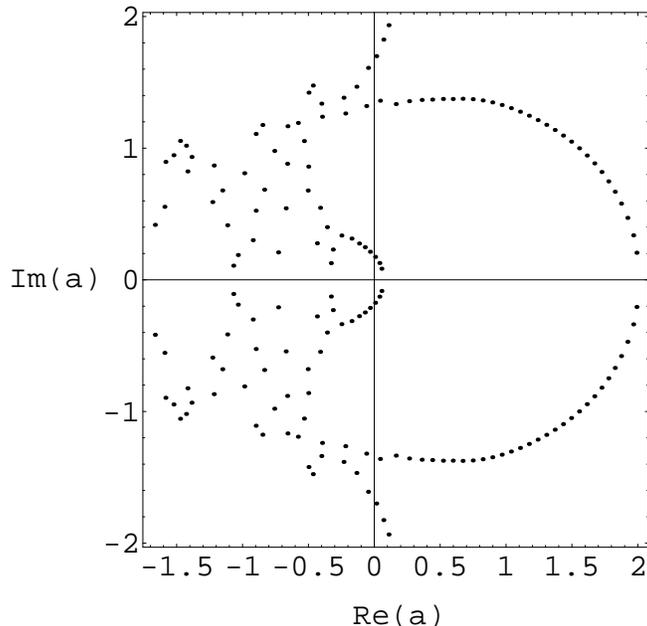}
\caption{CT zeros of $Z$ for the $q=4$ Potts model on the triangular lattice,
obtained via duality from a honeycomb lattice of size
$8 \times 6$ hexagons and (pbc,fbc)$_\Delta$ boundary conditions.}
\label{q4hexpfSpecaDual}
\end{figure}

In Figs. \ref{q4hexpfaDual} and \ref{q4hexpfSpecaDual} we show CT zeros of the
partition function for the $q=4$ Potts model on the triangular lattice with two
different sets of boundary conditions.  Qualitatively, the patterns of CT zeros
are similar to those for $q=3$.  The points at which the zeros, and hence the
CT phase boundary ${\cal B}$ inferred in the thermodynamic limit, cross the
real $a$ axis can be obtained via duality from those for the $q=4$ model on the
honeycomb lattice.  The rightmost crossing is consistent with the known 
exact value $a_{t,PM-FM,q=4}=2$ in eq. (\ref{at1q4}).  This is equivalent, by
duality, to the known value $a_{hc,PM-FM,q=4}=5$ for the $q=4$ Potts model on 
the honeycomb lattice.  Assuming the correctness of the suggested
zero-temperature critical point in the model \cite{baxter}, we deduce that the
CT zeros and the CT phase boundary ${\cal B}$ have no further crossings on the
positive real axis but cross this axis at $a=0$.  This would imply that in the
thermodynamic limit, the innermost complex-conjugate arcs of zeros pinch
together at this point.  As can be seen from Figs. \ref{q4hexpfaDual} and 
\ref{q4hexpfSpecaDual}, this requires that the two complex-conjugate arcs of
zeros nearest to the origin in the $a$ plane must pull back slightly to the
left as the lattice size goes to infinity.  By duality, this crossing of the CT
zeros at $a=0$ on the triangular lattice is equivalent to the crossing of the
CT zeros at $a=-3$ for the $q=4$ Potts model on the honeycomb lattice. 
In the present case, the CT zeros are also consistent with the conclusion 
that another crossing is at $a=a_{t,2,q=4}=-1$, the multiple root of eq. 
(\ref{trieqa}) given above in eq. (\ref{at2q4}); the dual equivalent is that
the CT zeros in the $q=4$ Potts model on the honeycomb lattice also cross the 
real $a$ axis at $a=-1$.  This is, of course, consistent with our calculations
of CT zeros on the honeycomb lattice.  There are also suggestions of other
possible crossings of CT zeros for the present $q=4$ case on the triangular
lattice.  These include a possible crossing at $a=-0.33(3)$, corresponding to 
the observed crossing of zeros for the honeycomb lattice at $a=-2.0(2)$.  We
have noted above that an analysis of the low-temperature series for the $q=4$
Potts model on the triangular lattice does not yield a firm indication of a 
singularity on the negative real $z$ (equivalently $a$) axis.  Such a
singularity would occur at the leftmost crossing of the zeros, $z_\ell =
a_{\ell}^{-1}$.  It is possible that the reason for this is that, if, indeed, 
$a_{t,\ell,q=4}=-1$, then the effect of this singularity is shielded by the
effects of singularities lying to the left in the complex $a$ plane (perhaps
associated with the arcs of zeros in Figs. \ref{q4hexpfaDual} and 
\ref{q4hexpfSpecaDual}), i.e., closer to the origin of the $z$ plane.  

  We also observe clear c. c. arcs of zeros protruding into the FM phase, with
endpoints at
\beq
a_{t,e,q=4}, \ a_{te,q=4}^* = 0.12(1) \pm 1.97(3)i
\label{aeq4}
\eeq
As indicated in the notation, this c. c. pair of points is in very good 
agreement with the c. c. pair of singularities given in eq. (\ref{aeq4ser}), 
identified from the analysis of the low-temperature series.  Thus, in both of
these cases, the $q=3$ and $q=4$ Potts models on the triangular lattice, we
find excellent agreement between such c. c. pairs of singularities extracted
from low-temperature series analyses and endpoints of arcs of zeros obtained
from the calculation of CT zeros on finite lattices.

\subsection{Comparison of CT Singularities with Phase Boundary for Triangular
Lattice}

   Evidently, the respective
values of the physical PM-FM critical point $a_{PM-FM,q}$
and the leftmost point where ${\cal B}$ crosses the real $a$ axis,
$a_{\ell,q}$, as obtained from the analysis of the low-temperature series are
in excellent agreement with the roots of the general formula (\ref{trieqa}) and
also with the crossing points seen with the CT zeros for both $q=3$ and $q=4$.
The respective
complex-conjugate pair of singularities at $z_{e,q},z_{e,q}^*$
from the analysis of the low-temperature series are seen to be the ends of
c. c. arcs of zeros (in the thermodynamic limit, continuous arcs of
singularities) protruding into and ending in the CT FM phase.  For $q=3$ and
$q=4$, from our analysis of the series we have also obtained
evidence for a complex-conjugate pair of singularities at the respective values
$a_{e',q},a_{e',q}^*$ as given in eqs. (\ref{afq3ser}) and (\ref{afq4ser}).
Comparing these respective pairs of singularities with the CT phase boundary
${\cal B}$ inferred from the CT zeros for $q=3$ and $q=4$, we observe that
each pair is consistent with lying on the respective boundary ${\cal B}$, in
the ``northwest'' and ``southwest'' quadrants of the $a$ plane.

\subsection{Triangular Lattice, $q=5$}

\begin{figure}
\centering
\
\epsfxsize=8.5 cm
\epsfbox{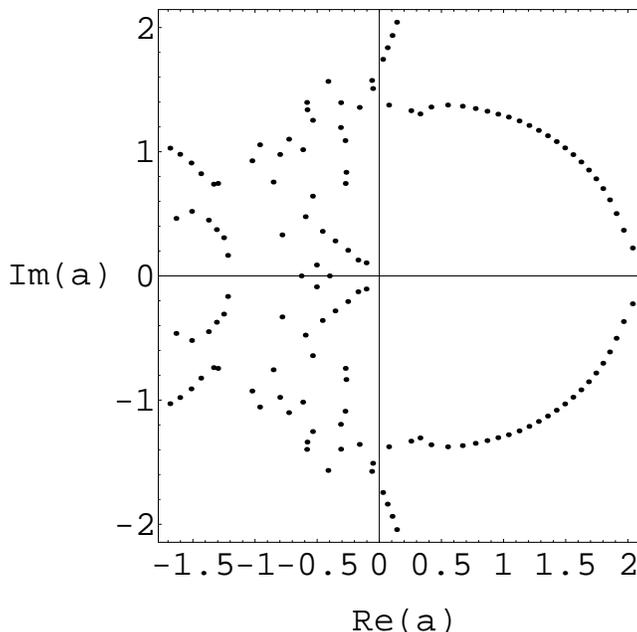}
\caption{CT zeros of $Z$ for the $q=5$ Potts model on the triangular lattice,
obtained via duality from a honeycomb lattice of size
$7 \times 6$ hexagons and (fbc,pbc) boundary conditions.}
\label{q5triangular}
\end{figure}

   It is also of interest to present an example of CT zeros calculated for the
$q$-state Potts model on the triangular lattice with $q$ in the range where the
Potts antiferromagnet is disordered at all temperatures including $T=0$.
Accordingly, we show in Fig. \ref{q5triangular} the CT zeros for the case
$q=5$.  For $q > 4$, eq. (\ref{trieqa}) has only one real root, together with a
conjugate pair of complex roots.  The real root for this case is
$a_{PM-FM,q=5}=2.103803...$.  This is in agreement with the rightmost crossing
point of the zeros on the real $a$ axis.  As is evident from
Fig. \ref{q5triangular}, the curves of zeros that we had inferred to pass
through the origin for the $q=4$ model have moved further to the left,
consistent with the conclusion that no branch of ${\cal B}$ passes through the
interval $0 \le a < 1$, i.e., that the $q=5$ Potts antiferromagnet is
disordered for all temperatures including $T=0$.  The other features of the CT
phase diagram are similar to those that we have observed before, including the
prominent c. c. arcs of zeros protruding into the FM phase near to the vertical
axis in the $a$ plane. 

\subsection{Comparison with $q=2$ Ising Case for the Triangular Lattice}

   One can gain some further insight into these results from a comparison with
the exactly solved $q=2$ Ising case.  The CT phase diagram in the $z$ plane is
shown as Fig. 1(b) of Ref. \cite{chitri}; in the $a$ plane, the CT phase
boundary maps to the union of (i) an oval with its longer side along the real
$a$ axis, crossing this axis at $\pm a_{PM-FM,q=2}=\pm \sqrt{3}$, and
(ii) a vertical line segment along the imaginary $a$ axis extending from
$-\sqrt{3}i$ upward to $\sqrt{3}i$.  This line segment bisects the oval and
divides the interior into the PM phase to the right and an O phase to the left;
outside of the oval is the FM phase.  The two components (i) and (ii) of
${\cal B}$ intersect at the multiple points $a=\pm i$.

   Now taking the broader perspective of general $q$, one sees that as $q$
increases from 2 to 4, the PM-FM phase transition point $a_{PM-FM,q}$, which is
the largest root of eq. (\ref{trieqa}), moves monotonically to the right, as
dictated by general inequalities (as $q$ increases, the spins become
``floppier'', and one must go to lower temperature to attain FM long-range
order).  A qualitative change occurs as one increases $q$ above 2, in that the
middle root of eq. (\ref{trieqa}) moves to negative values, as does this
portion of the CT phase boundary.  Thus, while the $T=0$ critical point of the
Ising antiferromagnet corresponds to the middle root of eq. (\ref{trieqa}), the
$T=0$ critical point of the $q=4$ model that has been argued for does not
correspond to any root of this eq. (\ref{trieqa}).

\section{Comparison of CT Singularities with Phase Boundaries for Other Cases}

   In Ref. \cite{p}, it was noted that for the Potts model with $q=3$ on the
honeycomb lattice and with $q=3,4$ on the kagom\'e lattice, the positions of
the physical PM-FM transition points, as obtained from eq. (\ref{hceqa}) for
the honeycomb lattice and from series analyses for the kagom\'e lattice, agreed
nicely with the maximal real points at which the CT zeros crossed the real $a$
axis.  It was also noted that the leftmost crossing point of the zeros at the
respective points $a_\ell$ were in good agreement (i) for the $q=3$ 
triangular case with a prediction from duality \cite{hcl} and a precise 
Monte Carlo measurement of the PM-AFM transition temperature of the $q=3$ 
Potts AF on the triangular lattice \cite{adler}; and (ii) for the $q=3,4$ 
kagom\'e case with the values obtained from low-temperature series analyses 
\cite{jge}.  Here, we extend this comparison to the complex-$a$ singularities. 

\subsection{Honeycomb Lattice, $q=3$}

   The low-temperature series analysis of the $q=3$ Potts model on the
honeycomb lattice in Ref. \cite{jge} yielded evidence for 
a c. c. pair of singularities in the thermodynamic functions at 
$z_\pm = -0.06(2) \pm 0.47(3)i$.  The central values correspond to 
$a_\pm = -0.27 \pm 2.1i$. Comparing with the results of Ref. \cite{p},
one sees that these points lie slightly to the upper left and lower left of 
the curve of CT zeros in the ``northwest'' and ``southwest'' quadrants of the 
$a$ plane.  This c. c. pair may be associated with possible c. c. cusp-like 
structures in this vicinity, as is hinted at in Fig. 7 of Ref. \cite{p}. 

\subsection{Kagom\'e Lattice, $q=3$} 

In addition to the PM-FM critical point and the leftmost crossing point,
the low-temperature series analysis of Ref. \cite{jge} found evidence for
CT singularities at four c. c. pairs of points in the $z$ plane, viz., 
$z_{1,\pm}=0.38(2) \pm 0.24(2)i$,
$z_{2,\pm}=0.278(10) \pm 0.38(1)i$,
$z_{3,\pm}=-0.113(6) \pm 0.515(10)i$, and 
$z_{4,\pm}=-0.37(2) \pm 0.30(5)i$.  The central values correspond approximately
to the points 
$a_{1,\pm}=1.9 \pm  1.2i$, 
$a_{2,\pm}=1.25 \pm 1.7i$, 
$a_{3,\pm}=-0.41 \pm 1.85i$, and 
$a_{4,\pm}=-1.6  \pm 1.3i$. 
As discussed above, one expects
the true singularities of the thermodynamic functions to lie on the CT phase
boundaries, since these quantities are analytic functions of
complex temperature in the interiors of the CT phases.  However, as our
illustration with the exactly calculated magnetization of the $q=2$ Ising case
showed, low-temperature series may, in general, indicate 
singularities which lie off the CT phase boundary in regions where these 
series do not apply; the corresponding factors are presumably present in the
true function (e.g. the factor $(1-u)^{-1/2}$ in (\ref{m_ising})).   \
The poles labeled
$a_{2,\pm}$ and $a_{4,\pm}$ are definitely consistent with lying on CT phase 
boundaries which may be plausibly inferred in the thermodynamic limit from 
the zeros calculated in Ref. \cite{p} for finite lattices.  For some, but not
all, types of boundary conditions, there is an indication of c. c. arcs of
zeros protruding into the CT FM phase and ending therein at points near to 
$a_{4,\pm}$.  The previous experience with exactly known CT phase diagrams 
\cite{g75,chitri,cmo} and with a
comparison between CT zeros and low-temperature series expansions for the
higher-spin Ising model \cite{ghs,hs} showed that, in the cases studied, the
magnetization diverges at the ends of arcs or line segments of singularities of
the free energy which protrude into the CT FM phase.  Thus, if the c. c.
singularities at $a_{4,\pm}$ do lie at the ends of such arcs, this would be
in accord with the divergence found in the magnetization in Ref. \cite{jge}.
The points $a_{3,\pm}$ lie somewhat outside the curves of zeros, in the
interior of the FM phase, while the points $a_{1,\pm}$ lie slightly inside of
the CT phase boundary ${\cal B}$ in the interior of the CT PM phase.  

\subsection{Kagom\'e Lattice, $q=4$}

   In this case, Ref. \cite{jge} obtained the two c. c. pairs of CT
singularities at $z_{1,\pm}=0.275(10) \pm 0.305(10)i$ and 
$z_{2,\pm}=-0.345(10) \pm 0.235(1)i$.  The central values correspond to the
points $a_{1,\pm}=1.63 \pm 1.81i$, and $a_{2,\pm}=-1.98 \pm 1.35i$. 
scattered set of poles in the region of the second pair.) The c. c. pair 
$a_{1,\pm}$ lie on the inferred phase boundary ${\cal B}$ in the upper and 
lower right quadrants of the $a$ plane.  Similarly, the c. c. pair 
$a_{2,\pm}$ lie on ${\cal B}$ in the upper and lower quadrants of the left-hand
half plane $Re(a) < 0$, in the region of arc-like structures on this boundary. 

\section{Conclusions}

   We have carried out a unified study of the $q$-state Potts model with $q=4$
on the honeycomb lattice and with $q=3,4$ on the triangular lattice, 
including the calculation and analysis of long low-temperature
series and the calculation of complex-temperature zeros of the partition
function which allow one to make reasonable inferences about the CT phase
boundary ${\cal B}$ in the thermodynamic limit.   In all cases, the series
are in excellent agreement with the known values of the respective
PM-FM critical points.   For the $q=4$ Potts model on 
the honeycomb lattice, there is no PM-AFM critical point and, concerning CT 
properties, we find that the
series analysis and CT zeros yield a value of the leftmost crossing $a_\ell$ in
good agreement with the inference from duality and the zero-temperature
critical point of the $q=4$ Potts model on the triangular lattice, viz., 
$a_\ell=-3$.  For the triangular lattice, the CT
zeros agree well with the known PM-AFM transition of the $q=3$ model and are
also consistent with the property that the $q=4$ model has a $T=0$ critical
point.  The singularities seen in the series at the largest negative values of
$a$ are seen to be the leftmost points where the CT phase boundary crosses
the negative real $a$ axis.  For both $q=3$ and $q=4$ the series also yield
clear indications of a complex-conjugate pair of singularities which are seen
to lie at the ends of arcs of CT zeros protruding into the CT FM phase.  In
each case, there are indications of another c. c. pair lying on the
respective CT phase boundaries.
We have also discussed how the positions of various CT 
singularities lying at complex values of $a$ in this model and also in the 
$q=3,4$ model on the kagom\'e lattice correlate with the respective CT phase 
boundaries. 

\vspace{2mm}

\begin{center}
{\bf Acknowledgments} 
\end{center}

   Financial support from the Australian Research Council is gratefully
acknowledged by I.J. and A.J.G.  The research of H.F., R.S., and S.-H.T. was
partially supported by the U.S. National Science Foundation under the grant 
PHY-97-9722101, for which these authors also express gratitude. 

\eject

\begin{table}
\renewcommand{\arraystretch}{0.5}
\caption{\label{table:hc4ser} Low-temperature series for the
4-state honeycomb lattice Potts model
magnetization ($m(z) = \sum_{n}m_{n}z^{n}$), susceptibility
($\bar{\chi}(z) = \sum_{n}x_{n}z^{n}$), and specific heat
($\bar{C}(z) = \sum_{n}c_{n}z^{n}$).}
\begin{center}
\begin{tabular}{rrrr}
 $n$ & $m_{n}$ & $x_{n}$ & $c_{n}$ \\
\hline
0   &   1
&   0   &   0   \\
1   &   0
&   0   &   0   \\
2   &   0
&   0   &   0   \\
3   &   -4
&   6   &   54   \\
4   &   -12
&   36   &   144   \\
5   &   -60
&   234   &   900   \\
6   &   -220
&   1284   &   2916   \\
7   &   -936
&   6804   &   14112   \\
8   &   -4092
&   38160   &   59616   \\
9   &   -17840
&   198912   &   280908   \\
10   &   -80868
&   1070316   &   1304100   \\
11   &   -356172
&   5499054   &   5974254   \\
12   &   -1640872
&   29005692   &   28501416   \\
13   &   -7433604
&   149318838   &   133160508   \\
14   &   -34541160
&   776570508   &   641771424   \\
15   &   -159080304
&   3987307152   &   3037720320   \\
16   &   -743832276
&   20560750344   &   14671207872   \\
17   &   -3469487112
&   105345948384   &   70242548778   \\
18   &   -16321682424
&   540305120844   &   340125653664   \\
19   &   -76796957940
&   2761471319562   &   1640652533460   \\
20   &   -363235185312
&   14111436147228   &   7963315328520   \\
21   &   -1720415299660
&   71964766006350   &   38614602921930   \\
22   &   -8176521038556
&   366780011157360   &   187903674109404   \\
23   &   -38925659520072
&   1866864944056032   &   914552556040350   \\
24   &   -185771131129720
&   9495487987576116   &   4460734444147344   \\
25   &   -888069677637192
&   48251046682543824   &   21771823449345750   \\
26   &   -4253549708242236
&   245022903414632628   &   106415060736772476   \\
27   &   -20404302611163396
&   1243326018023082990   &   520535130747734844   \\
28   &   -98033976216116940
&   6305270741929760652   &   2548904536404499392   \\
29   &   -471655884252852348
&   31956599345155563546   &   12490681376369529306   \\
30   &   -2272238036173908576
&   161878582502746522164   &   61260473924462872080   \\
\end{tabular}
\end{center}
\end{table}

\eject

\begin{table}
\renewcommand{\arraystretch}{0.5}
\caption{\label{table:tri3ser} Low-temperature series for the
3-state triangular lattice Potts model
magnetization ($m(z) = \sum_{n}m_{n}z^{n}$), susceptibility
($\bar{\chi}(z) = \sum_{n}x_{n}z^{n}$), and specific heat
($\bar{C}(z) = \sum_{n}c_{n}z^{n}$).}
\squeezetable
\begin{center}
\begin{tabular}{rrrr}
 $n$ & $m_{n}$ & $x_{n}$ & $c_{n}$ \\
\hline
0   &   1
&   0   &   0   \\
1   &   0
&   0   &   0   \\
2   &   0
&   0   &   0   \\
3   &   0
&   0   &   0   \\
4   &   0
&   0   &   0   \\
5   &   0
&   0   &   0   \\
6   &   -3
&   2   &   72   \\
7   &   0
&   0   &   0   \\
8   &   0
&   0   &   0   \\
9   &   0
&   0   &   0   \\
10   &   -18
&   24   &   600   \\
11   &   -18
&   24   &   726   \\
12   &   24
&   -20   &   -1440   \\
13   &   0
&   0   &   0   \\
14   &   -171
&   366   &   7056   \\
15   &   -162
&   324   &   8100   \\
16   &   153
&   -42   &   -13824   \\
17   &   252
&   -312   &   -20808   \\
18   &   -1704
&   4788   &   94176   \\
19   &   -2106
&   6036   &   119130   \\
20   &   1998
&   -1356   &   -196800   \\
21   &   2586
&   -1820   &   -291942   \\
22   &   -14364
&   54036   &   917664   \\
23   &   -28098
&   99252   &   1986924   \\
24   &   19008
&   -3024   &   -2389248   \\
25   &   43020
&   -53352   &   -5092500   \\
26   &   -147024
&   686988   &   10788960   \\
27   &   -317304
&   1382336   &   26041338   \\
28   &   125775
&   285870   &   -21643104   \\
29   &   612954
&   -926172   &   -81270876   \\
30   &   -1370868
&   7988984   &   111771360   \\
31   &   -3909528
&   19975392   &   369058596   \\
32   &   907209
&   6245886   &   -215519232   \\
33   &   7487136
&   -12161464   &   -1109316384   \\
34   &   -11849868
&   89970804   &   975825840   \\
35   &   -46762686
&   273568968   &   5032861050   \\
36   &   252159
&   134393334   &   -1479323520   \\
37   &   95554296
&   -181279824   &   -15448628352   \\
38   &   -101751129
&   1023192774   &   7864780656   \\
39   &   -543365058
&   3619881892   &   65059375680   \\
40   &   -122514741
&   2436896022   &   -501168000   \\
41   &   1155684132
&   -2347049916   &   -204974863146   \\
42   &   -703522230
&   10960701972   &   30675861720   \\
43   &   -6365905992
&   47574029772   &   839928958800   \\
44   &   -2758467240
&   39732936192   &   205816597536   \\
45   &   13464222858
&   -27776348840   &   -2605531430700   \\
46   &   -2746064529
&   113242596582   &   -499814655264   \\
47   &   -73051066008
&   609802710144   &   10503247729086   \\
48   &   -49228732689
&   624311338494   &   5545501277184   \\
49   &   154702726236
&   -310099907604   &   -32529619836714   \\
50   &   27843506676
&   1118687211276   &   -16889519112000   \\
51   &   -824524729038
&   7680614520344   &   127594218106044   \\
52   &   -769717612998
&   9376180586412   &   106385351442240   \\
53   &   1712690965746
&   -2931391777128   &   -391772958832758   \\
54   &   1028360456820
&   10088397834056   &   -346803904520640   \\
55   &   -9179822752182
&   95352726717060   &   1514773303324380   \\
56   &   -11186857401165
&   136258369372986   &   1772276306524416   \\
57   &   18287963891184
&   -19982241659216   &   -4555690872068178   \\
58   &   19778864095701
&   79744569755022   &   -5942932736977104   \\
59   &   -99841772973294
&   1162198685059320   &   17371599040182528   \\
60   &   -155837562896784
&   1933440869909764   &   27467732378426400   \\
61   &   186952834687950
&   15265872471072   &   -51015725275014492   \\
62   &   315816183555867
&   459254636055438   &   -92730567932042472   \\
63   &   -1058267389015764
&   13930030719657636   &   191575464300372474   \\
64   &   -2095009390517868
&   26811413231763564   &   403304437878595584   \\
65   &   1783741344539292
&   4261574859846552   &   -541403899076919450   \\
66   &   4604880525574113
&   -285699911125030   &   -1366060254075157608   \\
67   &   -10852791490392174
&   164052498128398560   &   2008868679625758660   \\
68   &   -27404067162573072
&   364675626055119000   &   5689560499409542368   \\
69   &   15230158436520024
&  &   -5331645029087453988   \\
\end{tabular}
\end{center}
\end{table}

\eject

\begin{table}
\renewcommand{\arraystretch}{0.5}
\caption{\label{table:tri4ser} Low-temperature series for the
4-state triangular lattice Potts model
magnetization ($m(z) = \sum_{n}m_{n}z^{n}$), susceptibility
($\bar{\chi}(z) = \sum_{n}x_{n}z^{n}$), and specific heat
($\bar{C}(z) = \sum_{n}c_{n}z^{n}$).}
\squeezetable
\begin{center}
\begin{tabular}{rrrr}
 $n$ & $m_{n}$ & $x_{n}$ & $c_{n}$ \\
\hline
0   &   1
&   0   &   0   \\
1   &   0
&   0   &   0   \\
2   &   0
&   0   &   0   \\
3   &   0
&   0   &   0   \\
4   &   0
&   0   &   0   \\
5   &   0
&   0   &   0   \\
6   &   -4
&   3   &   108   \\
7   &   0
&   0   &   0   \\
8   &   0
&   0   &   0   \\
9   &   0
&   0   &   0   \\
10   &   -24
&   36   &   900   \\
11   &   -48
&   72   &   2178   \\
12   &   60
&   -72   &   -3672   \\
13   &   0
&   0   &   0   \\
14   &   -300
&   711   &   14112   \\
15   &   -480
&   1080   &   27000   \\
16   &   144
&   144   &   -18432   \\
17   &   1392
&   -2556   &   -114444   \\
18   &   -4392
&   12852   &   290628   \\
19   &   -7248
&   23004   &   467856   \\
20   &   2904
&   -504   &   -354600   \\
21   &   13280
&   -21192   &   -1479114   \\
22   &   -27348
&   122877   &   1768536   \\
23   &   -142512
&   525996   &   12073896   \\
24   &   29948
&   69366   &   -5861808   \\
25   &   241872
&   -531576   &   -29193750   \\
26   &   -336072
&   1970154   &   22900176   \\
27   &   -1711936
&   7833756   &   165214728   \\
28   &   -950268
&   6613164   &   64153152   \\
29   &   4759680
&   -12953124   &   -654007014   \\
30   &   -2790212
&   24243261   &   163350540   \\
31   &   -25599600
&   137623572   &   2795893038   \\
32   &   -17648472
&   130318974   &   1579935744   \\
33   &   53777216
&   -138059232   &   -8670448116   \\
34   &   24551472
&   115953372   &   -6567077628   \\
35   &   -385317888
&   2338653528   &   48371018850   \\
36   &   -379526360
&   2854976280   &   42263145144   \\
37   &   757341312
&   -2039815332   &   -136194436566   \\
38   &   678358092
&   433835991   &   -156222316800   \\
39   &   -4605291200
&   33211423320   &   620105092776   \\
40   &   -8295782520
&   62346454416   &   1141035505200   \\
41   &   9858368640
&   -23201806140   &   -2037399494436   \\
42   &   16985972056
&   -32015102700   &   -3672631070136   \\
43   &   -59595025824
&   499277941344   &   8580306247938   \\
44   &   -136999873260
&   1118920518738   &   21031881323904   \\
45   &   81105525424
&   49166169540   &   -21085184576700   \\
46   &   365702657748
&   -1181015194617   &   -81272696605524   \\
47   &   -702809980704
&   6981782741964   &   105969879959940   \\
48   &   -2310592067252
&   20279205891438   &   391533801047712   \\
49   &   494570429328
&   6051532683060   &   -199750003201224   \\
50   &   5981082924792
&   -22190276278656   &   -1418750008893000   \\
51   &   -6198365886016
&   84541391331996   &   823212895965966   \\
52   &   -37809130736064
&   354875036323788   &   7044054404212800   \\
53   &   -6090856346112
&   205839844753932   &   -409244046301098   \\
54   &   97187254024404
&   -407434833461367   &   -24496576650092484   \\
55   &   -39928634332608
&   984502454280336   &   1572414830767440   \\
56   &   -562081834061556
&   5754184442187300   &   112350099811614336   \\
57   &   -340608212779056
&   5393530501373556   &   50948011956216330   \\
58   &   1482012248480712
&   -6542615994118038   &   -401080749409250964   \\
59   &   283544734049040
&   8894396469832452   &   -181847403757220400   \\
60   &   -8231813619904556
&  &   1752211155816298440   \\
\end{tabular}
\end{center}
\end{table}

\eject


\begin{thebibliography}{99}
\bibitem{com}{The free energy also has an isolated singularity at 
$|K|=\infty$ and (see Theorem 6 of Ref. \cite{cmo}) for the Ising model on 
lattices with odd coordination number, a singularity at $z=-1$.  The latter 
lies on the complex-temperature phase boundary for the honeycomb lattice but is
isolated for the heteropolygonal $3 \cdot 12^2$ lattice.}
\bibitem{cte}{As discussed in Ref. \cite{chisq}, the complex-temperature
extension of a physical phase is obtained by analytically continuing the free 
energy from the interval of physical temperature to a maximal region allowed by
nonanalytic boundaries.  Henceforth, we shall generally take the adjective 
``complex-temperature extension'' as implicit when referring to phases.  There
are also other complex-temperatures that have no overlap with any physical
phase; we shall denote these by O for ``other''.}
\end{thebibliography}
\end{document}